\documentclass[useAMS,usenatbib]{mn2e}

\usepackage{graphicx}
\usepackage{ulem}

\def\be{\begin{equation}}
\def\ee{\end{equation}}
\def\ba{\begin{eqnarray}}
\def\ea{\end{eqnarray}}

\def\msun{M_\odot}

\def\ltsima{$\; \buildrel < \over \sim \;$}
\def\simlt{\lower.5ex\hbox{\ltsima}}
\def\gtsima{$\; \buildrel > \over \sim \;$}
\def\simgt{\lower.5ex\hbox{\gtsima}}
\def\etal{{et al.\ }}

\def\zsun{Z_\odot}

\usepackage[dvipsnames]{color}

\title[OVI haloes]
{Extended OVI haloes of starforming galaxies}
\author[E. O. Vasiliev \etal]
       {Evgenii O. Vasiliev$^{1,2}$\thanks{E-mail:eugstar@mail.ru},
        Marina V. Ryabova$^2$\thanks{E-mail:mryabova@sfedu.ru},
        Yuri A. Shchekinov$^{2,3}$\thanks{E-mail:yus@sfedu.ru}\\
$^1$Institute of Physics, Southern Federal University, Stachki Ave. 194, Rostov-on-Don, 344090 Russia\\
$^2$Department of Physics, Southern Federal University, Sorge Str. 5, Rostov-on-Don, 344090 Russia\\
$^3$Isaac Newton Institute of Chile, SAO Branch\\
}
\begin{document}
\date{Accepted 3004 December 15.
      Received 2004 December 14;
      in original form 2004 December 31}
\pagerange{\pageref{firstpage}--\pageref{lastpage}}
\pubyear{3004}
\maketitle

\label{firstpage}

\begin{abstract}
We consider evolution of metal-enriched gas exposed to a superposition of time-dependent radiation field 
of a nearby starburst galaxy and nearly invariant (on timescales 100 Myr) extragalactic ionization background. 
Within nonequilibrium (time-dependent) photoionization models we determine ionization fraction of the OVI ion 
commonly observed in galactic circumference. We derive then conditions for OVI to appear in absorptions in 
extended galactic haloes depending on the galactic mass and star formation rate. We have found that the maximum 
OVI fraction can reach $\sim 0.4-0.9$ under combined action of the galactic and the extragalactic ionizing radiation 
fields. We conclude that soft X-ray emission with $E\simgt 113$~eV from the stellar population of central starforming 
galaxies is the main source of such a high fraction of OVI. This circumstance can explain high column densities 
${\rm  N(OVI)} \sim 10^{14.5 - 15.3}$~cm$^{-2}$ observed in the haloes of starforming galaxies at low redshifts
\citep{tumlinson11} {\it even} for a relatively low ({  $\sim 0.01-0.1\zsun$}) metallicity. As a result, the 
requirements to the sources of oxygen in the extended haloes relax to a reasonably conservative level. We show 
that at $z\simlt 0.5$ ionization kinetics of oxygen in a relatively dense plasma $n\simgt 10^{-4}$ cm$^{-3}$ of 
outer halo exposed to a low extragalactic ionizing flux is dominated by nonequilibrium effects.
\end{abstract}



\section{Introduction}

\noindent

Strong OVI absorptions observed around starforming galaxies at $z\sim 0.1-0.4$ with impact parameter as high as 150 
kpc reveal huge gaseous galactic haloes \citep{tumlinson11}. Even conservative estimates lead to amount 
of gas in them far exceeding the gas reservoir in galaxies themselves. Such a conclusion is based on the assumption 
that gas in the haloes has solar metallicity. The arguments underlying this assumption stem from the standard estimate 
of oxygen mass from the observed column density \citep{tumlinson11}
\be
\label{mo}
M_{\rm O} = 5\pi R^2\langle N_{_{\rm OVI}}\rangle m_{_{\rm O}} f_{hit} \left( {0.2 \over f_{_{\rm OVI}}} \right)
          = 1.2\times 10^7 \left({0.2\over f_{_{\rm OVI}}}\right)M_\odot,
\ee
where a typical column density $\langle N_{_{\rm OVI}}\rangle=3\times 10^{14}$ cm$^{-2}$, the halo radius $R=150$ kpc, 
the hit rate correction factor {  (covering factor) $f_{hit}=0.8-1$} are assumed following by \citet{tumlinson11}, 
{  $m_{_{\rm O}} = 16m_{_{\rm H}}$ is atomic mass of oxygen. }
The fractional abundance of OVI under thermal collisional ionization equilibrium never exceeds $f_{_{\rm OVI}}=0.2$
\citep[e.g.][]{cloudy,gs07}, such that within this assumption Eq. (\ref{mo}) provides a lower estimate of oxygen mass in the
halo \citep{tumlinson11}. It is readily seen from here that the mass $M_{\rm O}\sim 10^7M_\odot$, i.e. around
10 to 70\% of oxygen mass in the ISM, is indeed a conservative estimate.

In principle, this conclusion for galactic haloes to bear such a large gas mass would might solve the problem 
of missing baryons and metals \citep{bregman,pettini}, though requiring enormously high oxygen production and mass 
ejection rates. Moreover, the fraction $f_{_{\rm OVI}}=0.2$ under collisional ionization equilibrium is kept 
only in a very narrow temperature range: $T=(3-5)\times 10^5$ K \citep[e.g.][]{cloudy,gs07}. It is {  therefore} 
totally unrealistic to assume that all observed haloes keep their temperatures within {  such a narrow range independent} 
on distance from the host galaxy.

In their estimates \citet{tumlinson11} used the ionic fractions {  calculated} under collisional and/or photoionization
equilibrium {  conditions}, i.e. independent of time \citep{cloudy}. However, ionization state of gas situated in
time-dependent (nonequilibrium) environment can differ qualitatively from the one {  settled on to} equilibrium,
particularly, for solar metallicity \citep{v11}. This difference can be smaller for low-density gas exposed to a strong
ionizing field. However, in the process of galactic evolution both the magnitude and the shape of the radiation spectrum can change. 
{  In addition the} extragalactic spectrum, which can be important on the periphery of galactic haloes, does also evolve
\citep{hm01,faucher}.  Under such conditions it is natural to expect  the ionic composition {to  experience  
time variations.}

In this paper we therefore concentrate on the question of whether evolution of the ionizing radiation field
can result in considerable changes of fractional ionization of oxygen such to make estimates of gas mass in galactic 
haloes more reliable. In next sections we will demonstrate that indeed under time-dependent conditions the observed 
column densities of OVI correspond to at least half order of magnitude less massive haloes, which can stay in a much 
wider range of physical parameters.

The paper is organized as follows. In Section 2 we describe the details of the model. In Section 3 we present our
results. Section 4 summarizes the results.


\section{Model description}

\noindent
Thermal and ionization state in our model is fully time-dependent: the model involves ionization and thermal
evolution of gas located at radii $\sim 50-300$~kpc in the galactic halo, exposed to extragalactic
and galactic time-dependent ionizing radiation field. 

\subsection{Time-dependent ionization}

In this paper we only briefly touch main features of calculation of the ionization and thermal evolution
of gas immersed into external time-dependent ionizing radiation. The details can be found in \citep{v11}.
We study ionization and thermal evolution of a lagrangian gas element: 
a gas parcel is assumed to be optically thin to external ionizing radiation. In the  
calculations we include all ionization states of the elements H, He, C, N, O, Ne, Mg, 
Si and Fe. We take into account the following major processes: photoionization, 
multi-electron Auger ionization process, collisional ionization, radiative and 
dielectronic recombination, {  collisional excitation} as well as charge transfer 
in collisions with hydrogen and helium atoms and ions. 

The total cooling and heating rates are calculated using as a subroutine the photoionization 
code CLOUDY \citep[ver. 10.00,][]{cloudy}. More specifically, we input into CLOUDY code a given 
set of all ionic fractions $X_i$ calculated at temperature $T$, gas density $n$ and 
external ionization flux $J(\nu)$ and obtain the corresponding cooling and heating rates. 
{  The latter also includes Compton heating from X-rays.}
For the solar metallicity we adopt the abundances reported by \citet{asplund}, except 
Ne for which the enhanced abundance is adopted \citep{drake}. In all calculations we 
assume the helium mass fraction $Y_{\rm He} = 0.24$, which corresponds to [He/H]=0.081,  
and is close to the observed one \citep{izotov}.

We solve a set of 96 coupled equations (95 for ionization states and one for temperature)
using a Variable-coefficient Ordinary Differential Equation solver \citep{dvode}. 
We considered the two regimes of gas evolution: isochoric and isobaric. The isochoric 
regime suggests that gas density in a cloud is kept constant, while in isobaric model gas pressure 
is assumed constant. The two regimes correspond to two opposite limits of the ratio between the cooling and 
the sound crossing times: $t_{_c}=kT/\Lambda n$ and $t_{_s}=R/c_s$, correspondingly. In the external heating 
radiation field isobaric models   
show essentially similar thermal and ionization evolution, though on longer time scales due to 
decreasing density coupled to increasing temperature. It results in an increase of the cross-section 
of gas clouds such that their covering factor increases as well.

\subsection{Galaxy evolution}

In the process of galaxy evolution the stellar content changes: massive stars produce enormous number of UV 
photons and form ultimately compact objects which emit hard ionizing photons. Spectrum of the ionizing radiation 
escaping galactic interstellar medium (ISM) and exposing further the halo depends on amount of metals in the ISM 
disk absorbing ionizing photons, and thus on chemical evolution of the galaxy. In order to follow evolution of 
stellar mass, metallicity and galaxy spectrum, we use the spectro-photometric code PEGASE \citep{pegase97}. We 
assume a Schmidt-like power-law starformation rate (SFR): ${\rm SFR}(t) = {\cal M}_g^{p_1}/p_2$, typical SFR for 
massive starforming galaxies, where {  ${\cal M}$ is the normalized mass of gas in $\msun$}. In some regions of 
the galaxy SFR can be 
inhibited by gas outflows from the disk, however, when averaged over the whole disk the SFR remains sufficiently 
high over the whole period of active star formation. In our models we assumed a closed-box regime. In general, 
though this cannot be applied to galaxies with active star formation. However, many parameters related to mass 
and energy exchange between galaxies and the intergalactic medium, such as the rates of mass ejection from and 
mass accretion from the ambient medium, gas metallicity and corresponding cooling rate, clustering of SNe 
explosions, etc are highly uncertain and hard to be coherently described phenomenologically. 

\subsection{Time-dependent UV/X-ray backgrounds}

Gas in galactic haloes is exposed by a cumulative ionizing background consisting of the extragalactic and the galactic components. The extragalactic component is uniform on galactic halo scales, and is nearly constant on timescales $\sim
100$~Myr, while changes significantly on longer times. For the extragalactic background we accept the spectrum described 
by \citet{hm01}. Its evolution covers redshifts $z=0$ to $z=9$ divided by 49 equally spaced log bins. Contribution from
possible flux sources varies over cosmological time, and irregular changes in different bands of the spectrum with
redshift can be met. For this reason simple linear approximation between neighbour redshift bins was used.

The galactic component instead, may change on much shorter timescales. In general, the corresponding time scale is 
close to the lifetime of massive stars, i.e. $\sim 10-20$~Myr. This galactic radiation component originates in the 
central star forming region, and is seen from outer parts of the halo at distances $\simgt 30-50$~kpc as a nearly 
spherical domain of size of $\sim 2-3$~kpc.

The ionizing radiation {  from stellar population} is partly modified by absorption in the interstellar 
{  medium} of underlying bulge and 
disk. In order to account this absorption we assume that ionizing photons pass through {  a layer of neutral gas in 
galactic disc} with the 
optical depth $\tau_{\nu} = \sigma_\nu^{\rm HI}N_{\rm HI} + \sigma_\nu^{\rm HeI}N_{\rm HeI}$, throughout the 
paper $N_{\rm HI} =
10^{20}$~cm$^{-2}$, $N_{\rm HeI} = 10^{19}$~cm$^{-2}$ {  are considered as fiducial. The corresponding optical depths 
at the HI and HeI Lyman limits are as high as $\sim 630$ and $70$, respectively. As a result only photons with $E\simgt
60$~eV escape the galaxy and penetrate into halo; {  the ionizing flux with photons of $E>60$~eV decreases} 
as $r^{-2}$. In what follows we will discuss the dependence of our results on the HI column density. }

In our model the galactic UV spectrum is calculated with making use the PEGASE code \citep{pegase97}, which gives
spectral luminosity in the range from 91\AA \ to 160~$\mu$m. In order to extend the spectrum to higher energies 
(up to $\sim 10^4$ eV, responsible for ionization of highly charged ionic species) we use the empirical relation 
between the X-ray luminosity and the star formation rate `$L_X - SFR$' 
\citep{gilfanov04}. This relation is well established for massive starforming galaxies considered here. 

Overall the cumulative spectrum varies on time scales from several to hundreds of millions years. 

\subsection{Initial set up}

We consider gas in outer haloes of massive (Milky Way type) starforming galaxies with stellar mass of several 
$\times 10^{10}~\msun$. Recent simulations of the Milky Way halo show that it extends up to the virial radius 
of the Milky Way (i.e., $\sim 50-300$~kpc) with densities ranging within $\sim (0.5-2)\times 10^{-4}$~cm$^{-3}$
\citep{feldmann13}. Observational estimates of the circumgalactic gas density around the Milky Way and other 
Local Group galaxies give  similar numbers $\sim (1-3)\times 10^{-4}$~cm$^{-3}$ at $r\sim 40-150$~kpc
\citep{weiner96,grcevich09,quilis01,stanimirovic02,anderson10}. in our calculations we follow these numbers and 
set $n = (0.5-2)\times 10^{-4}$~cm$^{-3}$ in the circumgalactic volume. We consider both isochoric and isobaric 
regimes.

We start the calculations at $z=2$ (the lookback time is around 10 Gyrs). This timescale is nearly the cooling 
time for hot gas with $T\sim 10^6$~K and $\sim (0.5-2)\times 10^{-4}$~cm$^{-3}$ \citep{feldmann13}. The last 
major merging for Milky Way-type galaxies is thought to occur earlier than $z\sim 2$ \citep[e.g., ][]{mwmerger}. 

The initial ionic composition and the temperature are set equal to the ones corresponding to photoequilibrium in 
gas exposed to the extragalactic Haardt \& Madau spectrum at $z=2$. Such radiation field is sufficient to force 
low density gas with $n$ in the accepted range to settle quickly onto photoequilibrium \citep{v11}. Calculations 
cover physical time scales much longer than the relaxation time scale of ionization and thermal processes in gas 
exposed to the time-dependent spectrum adopted here.

Gas metallicity in our models is assumed to range within $10^{-2}$ to $0.1~\zsun$. Higher metallicities are rare 
in DLA QSO and DLA GRB absorbers \citep{sava09}. The lower limit lies around ten times above the upper limit of 
the IGM metallicity at $z\sim 2-3$ \citep[e.g.,][]{cowie95,dodorico10}.


\section{Results}

\subsection{SFR and spectral evolution}

\begin{figure}
\center
\includegraphics[width=80mm]{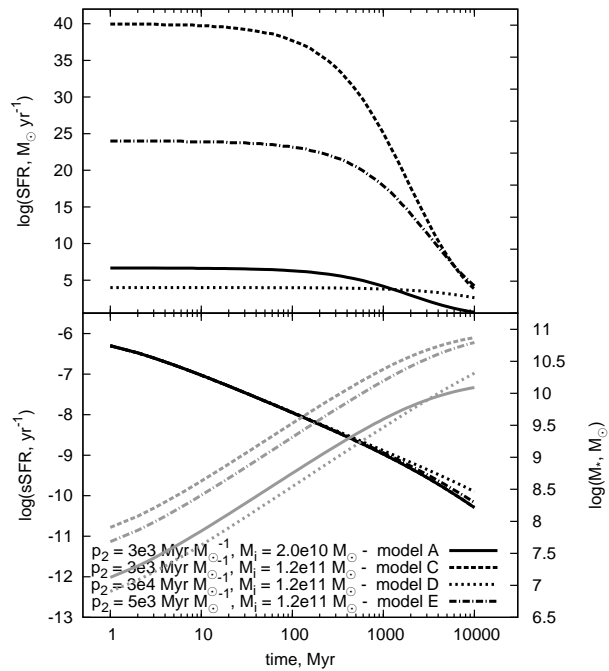}
\caption{
{  {\it Upper panel}: the star formation rate. {\it Lower panel}: the specific 
star formation rate, ${\rm sSFR} = {\rm SFR}/M_*$ (solid black),  
and the stellar mass, $M_*$ (solid grey);}   
{  the initial time corresponds to $z=2$.
The four models with the rate {  ${\rm SFR}(t) = {\cal M}^{2}/p_2$}
are shown: two models with fixed $p_2=3\times 10^3$~Myr~$\msun^{-1}$ and the initial gaseous mass 
$M_g^i = 2\times 10^{10}~\msun$ (model A -- solid line), and $M_g^i = 1.2\times 10^{11}~\msun$ (model C -- dashed line). 
{  The other two models are } with fixed $M_g^i = 1.2\times 10^{11}~\msun$ and different  
$p_2=3\times 10^4$~Myr~$\msun^{-1}$  (model D -- dotted line),  
 and $p_2=5\times 10^3$~Myr~$\msun^{-1}$ (model E -- dash-dotted line).
}
}     
\label{figssfr-time}
\end{figure}

From the chemical evolution models calculated with the PEGASE code we obtain the time-dependent starformation rate, 
stellar mass, gas metallicity and spectral luminosity. We simulated several models corresponding to massive galaxies 
and chose two of them, whose parameters, namely the SFR and the stellar mass, are close to those of the starforming
galaxies described by \citet{tumlinson11}. With the accepted SFR law ${\rm SFR}(t) = {\cal M}^{p_1}/p_2$, where 
{  ${\cal M}$ is the normalized mass of gas in $\msun$ and} through the paper $p_1 = 2$ is assumed, the notations are as 
in \citep{pegase97}. In total 
we consider four models. The first two suggest a fixed normalization factor $p_2=3\times 10^3$~Myr~$\msun^{-1}$, 
and {  two different values of the initial gaseous mass:} $M_g^i = 2\times 10^{10}~\msun$ -- model A, and $M_g^i = 1.2\times
10^{11}~\msun$ -- model C. The other two models assume {   a fixed initial gaseous mass} $M_g^i = 1.2\times 10^{11}~\msun$, 
and two different normalization 
constants $p_2$:  one with $p_2=3\times 10^4$~Myr~$\msun^{-1}$ -- model D, and the other with 
$p_2=5\times 10^3$~Myr~$\msun^{-1}$ -- model E.

Figure~\ref{figssfr-time} presents {  the star formation rate, $\rm SFR$}, the specific starformation rate, ${\rm sSFR} = {\rm
SFR}/M_*$, and the stellar mass, $M_*${: model A (solid line), model C (dashed line), model D (dotted line) and 
model E (dash-dotted line). 
Two features are to note: first, the sSFR reveals differences between the models only after several 
hundreds of Myrs, and second, models A and C show practically equal sSFR, and as a result gas in these models is equally  
converted into stars by $\sim 10$~Gyr. Instead, model D ($p_2=3\times 10^4$~Myr~$\msun^{-1}$) leaves the galaxies  
gas-rich -- only $\sim 20$~\% of gaseous mass exhausts by this time. The characteristic time for the gas to exhaust 
is $t_g\sim p_2/M_g$, which gives $t_g\simgt 300$ Myr for all considered models, resulting in a §nearly constant SFR 
(upper panel in Figure~\ref{figssfr-time}) over first 
200~Myr with the SFR$\propto p_2^{-1}$, and decreasing on later times as SFR$\propto p_2/t^2$. 
} 

\begin{figure}
\center
\includegraphics[width=80mm]{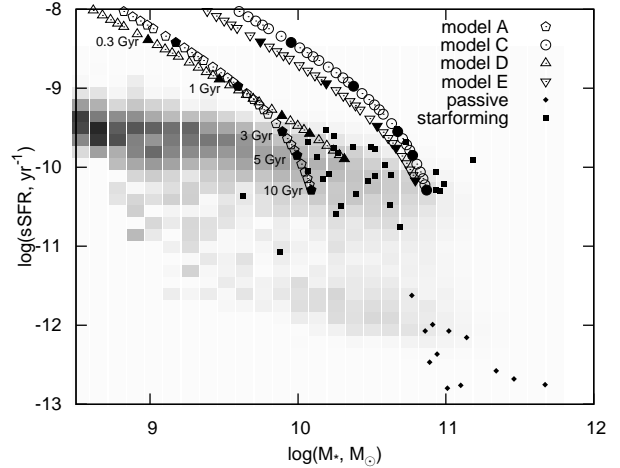}
\caption{
The dependence of the specific starformation rate on the stellar mass for models A {  (pentagons)}, C {  (circles),
D (up-triangles) and E (down-triangles)}.
Filled {  large symbols mark} time moments indicated in the Figure. Data for the starforming
and passive galaxies from \citet{tumlinson11} are depicted by {  small} squares and rhombs, correspondingly. 
The grey-scale map is for SDSS+GALEX galaxies \citep{schiminovich}. 
{  SFR models are described in text and Figure~\ref{figssfr-time}.}
}
\label{figsfr-mstar}
\end{figure}

Figure~\ref{figsfr-mstar} shows the dependence of specific starformation rate on the stellar mass for the models A
{  (pentagons)}, C {  (circles), D (up-triangles) and E (down-triangles)}. Filled {  large symbols 
mark} 
time moments shown nearby. Data for the starforming and passive galaxies studied by \citet{tumlinson11}
are depicted by {  small} squares and rhombs, correspondingly. The grey-scale map shows SDSS+GALEX galaxies
\citep{schiminovich}. It is  readily seen that almost all points for the starforming galaxies are locked between 
tracks of the models considered here. One can expect therefore that spectral properties of starforming galaxies from
\citet{tumlinson11} are similar to those in the models A and C during the latest $\sim$3-4~Gyrs of their evolution.
Passive galaxies have an order of magnitude lower sSFR. 

\begin{figure}
\includegraphics[width=80mm]{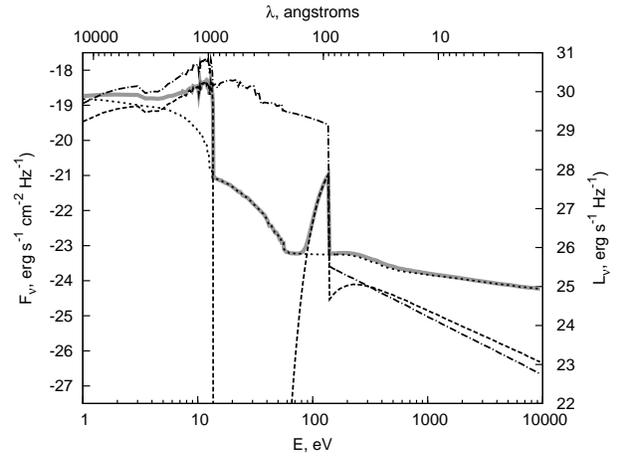}
\caption{
The cumulative ionizing background flux (thick grey line) at $z=0.2$ and a distance from the galaxy $d=100$~kpc, 
which consists of the UV and X-ray galactic spectrum, attenuated by galactic neutral gas (dash line), the 
extragalactic ionizing background (dotted line). The galactic spectral luminosity is shown by dash-dotted line 
(the right axis). 
}     
\label{figspec}
\end{figure}

Dash-doted line in Figure~\ref{figspec} shows the galactic spectral luminosity at $t=7.5$~Gyr, which is equal time 
elapsed from $z=2$ to 0.2. A significant break is clearly seen at the wavelength 91~\AA, which corresponds to the 
minimum wavelength in the spectrum reached in the PEGASE code\footnote{{  The cut at 91~\AA \ in the galactic spectral
energy distribution (SED) generated by the stellar population codes (e.g. PEGASE, Starburst99, Galaxev) is apparently due to 
a choice of the developers and might be partly connected with the availability of spectral data in stellar libraries.
The stellar library calculated by \citet{rauch03} contains spectra up to 1~\AA \ is not included yet into stellar 
population codes. It is worth also noting in this connection considerable deviations of the spectral distribution above
228~\AA \ predicted by different stellar population codes \citep{kewley01}. In this sense our conclusions may depend on 
our choice of the stellar population code {  and, particularly, on the cut at 91~\AA. In Section 3.3 we discuss stability 
of our results against possible variations of the galactic SED, and influence of the cut at 91~\AA \ on the ionization and 
thermal evolution of the circumgalactic gas.} }}. 
The break is due to an exponential decrease of the number of such hard photons emitted by stellar population \citep[see
e.g.][]{rauch03}, as they are only produced by very massive stars, whose number is very small. As soon as we consider galactic 
evolution on timescales longer than 1-3~Gyr X-ray binaries are expected to have already formed, and we extend the spectrum to 
higher energies assuming the empirical relation between X-ray luminosity and star-formation rate ``$L_X - SFR$'' \citep{gilfanov04}.

By the thick grey line Figure~\ref{figspec} shows also an example of the total spectral radiation flux exposing 
a given gas parcel located at distance $d=100$~kpc from the galaxy evolved till $z=0.2$. The total spectrum consists of 
the galactic (dash line) and extragalactic (dotted line) ionizing photons. At low energies, $E\simlt 13.6$~eV, 
the extragalactic contribution dominates at large distances, $d\simgt 100$~kpc, while the stellar population turns 
into play in UV range at smaller distances. Strong absorption of the galactic ionizing photons ($E\sim 13.6-90$~eV) 
in the galactic disk {  with the fiducial $N_{{\rm HI}}$ and $N_{{\rm HeI}}$} values results in the total dependence 
of the ionic composition of halo gas on the extragalactic background. The significance of absorption by the galactic 
disk can be understood from comparison of the galactic spectral luminosity shown by dash-dotted line and the cumulative 
flux (thick grey line).

A narrow bump at $E\sim 90-136$~eV is due to the galactic photons survived against absorption in thick disk. Its magnitude 
is obviously determined by our choice of the neutral column densities $N_{{\rm HI}}$ and $N_{{\rm HeI}}$ in the disk. For 
the fiducial values of $N_{{\rm HI}}$ and $N_{{\rm HeI}}$ the optical depth is about 3 for photons with $E\sim 113$~eV. 
Higher column densities can erase the bump and the extragalactic flux becomes dominating in the whole galactic halo. 
Decrease of the galactic contribution is also seen at larger distances from the galaxy due to dilution $r^{-2}$. Note that 
the ionization potential of OV, $I_{\rm OV} = 113.9$~eV falls exactly in this the range $E\sim 90-136$~eV. This means that 
the fraction of OVI ions can increase in galaxies with lower column densities ($N_{{\rm HI}}$ and $N_{{\rm HeI}}$) in their 
disks. 

\begin{figure}
\center
\includegraphics[width=80mm]{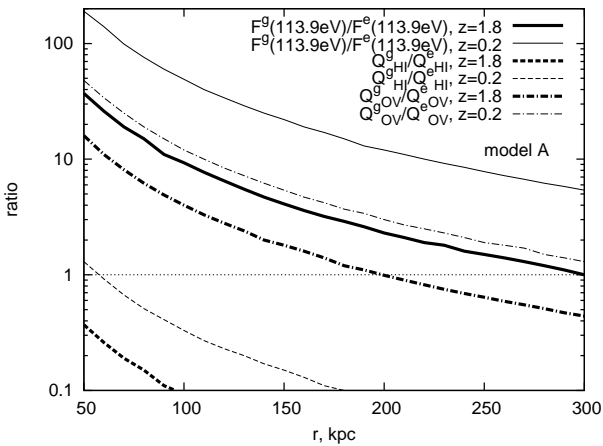}
\caption{
The ratio of monochromatic galactic flux to the extragalactic one at 113.9~eV (solid lines), and the ratios
of ionizing photon number fluxes $Q^g_{\rm HI}/Q^e_{\rm HI}$ (dash lines) and $Q^g_{\rm OV}/Q^e_{\rm OV}$ 
(dash-dotted lines), where the superscipts 'g' and 'e' refer to galactic and extragalactic backgrounds. 
Thick and thin lines correspond to redshifts 1.8 and 0.2, respectively. 
}
\label{figflux-ratio}
\end{figure}

In general, the excess of galactic photons with energies higher than 113.9~eV contributes crucially into ionization kinetics 
of OVI ion. Figure~\ref{figflux-ratio} presents the ratio of monochromatic galactic and extragalactic fluxes at 113.9~eV 
(solid lines) in model A for our fiducial column densities $N_{{\rm HI}}$ and $N_{{\rm HeI}}$. It is clearly seen that the
galactic flux dominates within $r\simlt 300$~kpc at redshifts $z>0.2$.  At lower redshifts the region of predominance 
of galactic flux widens mostly due to a steep decrease of the extragalactic flux in these epochs. The monochromatic radiation
at the HI Lyman limit is fully absorbed in the disk for the fiducial $N_{{\rm HI}}$ and $N_{{\rm HeI}}$ column densities.

In order to estimate both the escape of HI ionizing photons and the efficiency of OV ionization we calculate 
radial dependences of the ratios of ionizing 
fluxes $Q^g_{\rm HI}/Q^e_{\rm HI}$ and $Q^g_{\rm OV}/Q^e_{\rm OV}$, where the superscipts '{\it g}' and '{\it e}' 
refer to the galactic and extragalactic contributions. Figure~\ref{figflux-ratio} shows these 
ratios at $z=1.8$ and 0.2 for model A. In the energy range $E\simgt 113.9$~eV the galactic contribution dominates 
up to distance $r\simlt 200$~kpc at $z=1.8$, and extends even to $r\simlt 300$~kpc at $z=0.2$
due to strong decrease of the extragalactic background at low redshifts. It is obvious, that increase of ionizing flux 
enlarges radius of the zone of predominance of galactic ionizing photons: for instance, this zone 
increases from $\sim 200$~kpc in model A to $\sim 300$~kpc 
in model C at $z=1.8$. Decrease of absorption in the disk does also promote the zone of the galactic predominance to grow: 
it reaches $r \simgt 250$~kpc for $N_{{\rm HI}} \simlt 10^{19}$~cm$^{-2}$ even in model~A.

\subsection{Thermal and ionization evolution}

\begin{figure}
\center
\includegraphics[width=80mm]{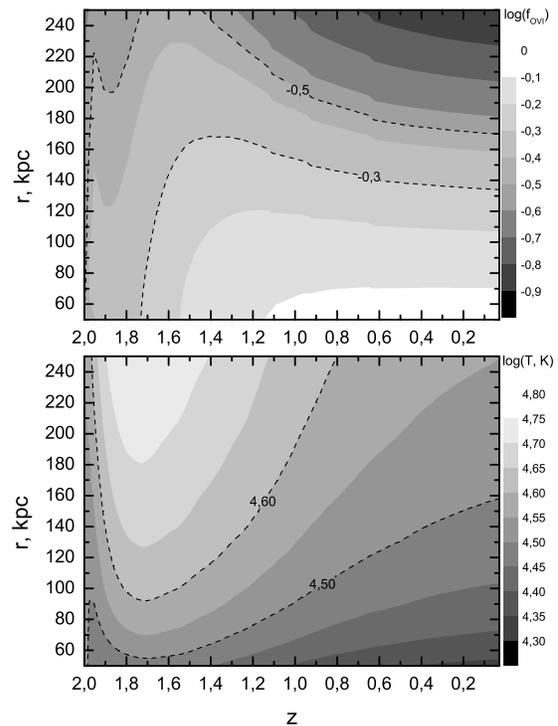}
\caption{
The temperature (lower panel) and OVI fraction (upper panel) evolution (dependence on redshift) of a gas at different
distances from the galactic center. {  We assume the SFR as that in model A and our fiducial values of $N_{{\rm HI}}$
and $N_{{\rm HeI}}$.} 
}
\label{figt-ovi-evol}
\end{figure}

Figure~\ref{figt-ovi-evol} shows evolution of temperature (lower panel) and OVI fraction (upper panel) in gas 
located at different distances from the galactic center; {  here we assume the SFR as in model A and our fiducial 
column densities $N_{{\rm HI}}$ and $N_{{\rm HeI}}$.}

In the beginning, $z\sim 2$, oxygen is mainly locked in the OVII state due to high ionizing flux.
Its fraction establishes around $\sim 0.5$ under the influence of the extragalactic ionizing background. After several
hundreds million years the extragalactic background starts decreasing following the cosmic star formation rate.  
It should result in a quick transition from OVII to OVI. In a cooling plasma OVI recombines rapidly into
lower ionization states and practically disappears shortly -- a well-known OVI ``fragility''. However, in our case 
the excess of photons with $E > I_{\rm OV} = 113.9$~eV emitted by starforming galaxies does not allow OVI to recombine.
Consequently, its fraction remains almost frozen at the level $\sim 0.5$ in the region $d\simlt 300$ kpc over the 
range $z\sim 1.2$ till 0. In our
models $f({\rm OVI})$ reaches $\sim 0.4-0.9$ in a low-metallicity ($0.1~\zsun$) gas within $50<d<150$~kpc from
$z\sim 1$ to $0$. This value is several times higher than the maximum OVI fraction, $\sim 0.1-0.2$, reached 
in gas exposed only to the extragalactic background \citep[e.g.][]{cloudy,gs07}, and in nonequilibrium collisional gas
evolved from $T=10^8$~K \citep[see e.g.][]{gs07}. Note that even at large distances $d\sim 250$ kpc $f({\rm OVI})$ 
remains higher than $0.1$. The temperature of gas with such high OVI fraction is within $(2-5)\times 10^4$~K (see 
lower panel in Figure~\ref{figt-ovi-evol}). The dependence of $f({\rm OVI})$ on metallicity is weak: for instance, 
OVI fraction in gas with $0.01~\zsun$ is $\sim 0.1-0.8$ within distances $50<d<300$~kpc in model~A.

Cooling of gas with $Z\simlt 0.1~\zsun$ exposed only to the extragalactic background is mainly due to hydrogen and
helium, whereas metals (oxygen and carbon) play a minor role in radiation losses \citep[e.g.][]{wiersma09,v11}. 
A considerable increase of OVI fraction in the zones where galactic ionization dominates enhances the contribution 
of oxygen into cooling. As a result, gas temperature is lower in these regions 
as seen on the lower panel in Figure~\ref{figt-ovi-evol}.

Deviation of column densities from the fiducial values accepted above may result in a considerable change of 
the overall picture due to changes of the interrelation between the fractions of galactic and extragalactic ionizing 
photons. In order to understand how sensitive is the oxygen ionization state to surface density of the underlying 
galactic gaseous discs we calculate several models with different $N_{\rm HI}$ and $N_{\rm He}$. 
For simplicity we assume that $N_{\rm HI} /N_{\rm HeI} =10$.

\begin{figure}
\center
\includegraphics[width=80mm]{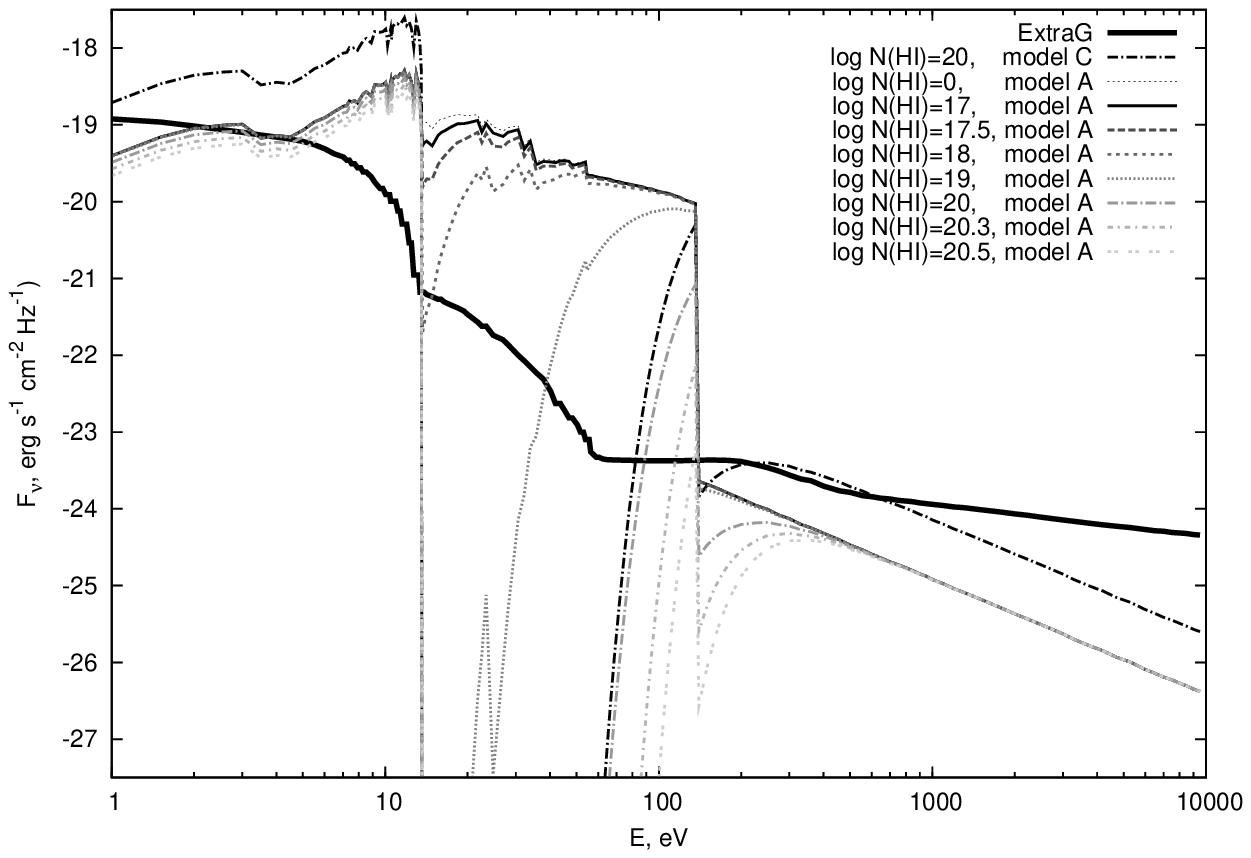}
\includegraphics[width=80mm]{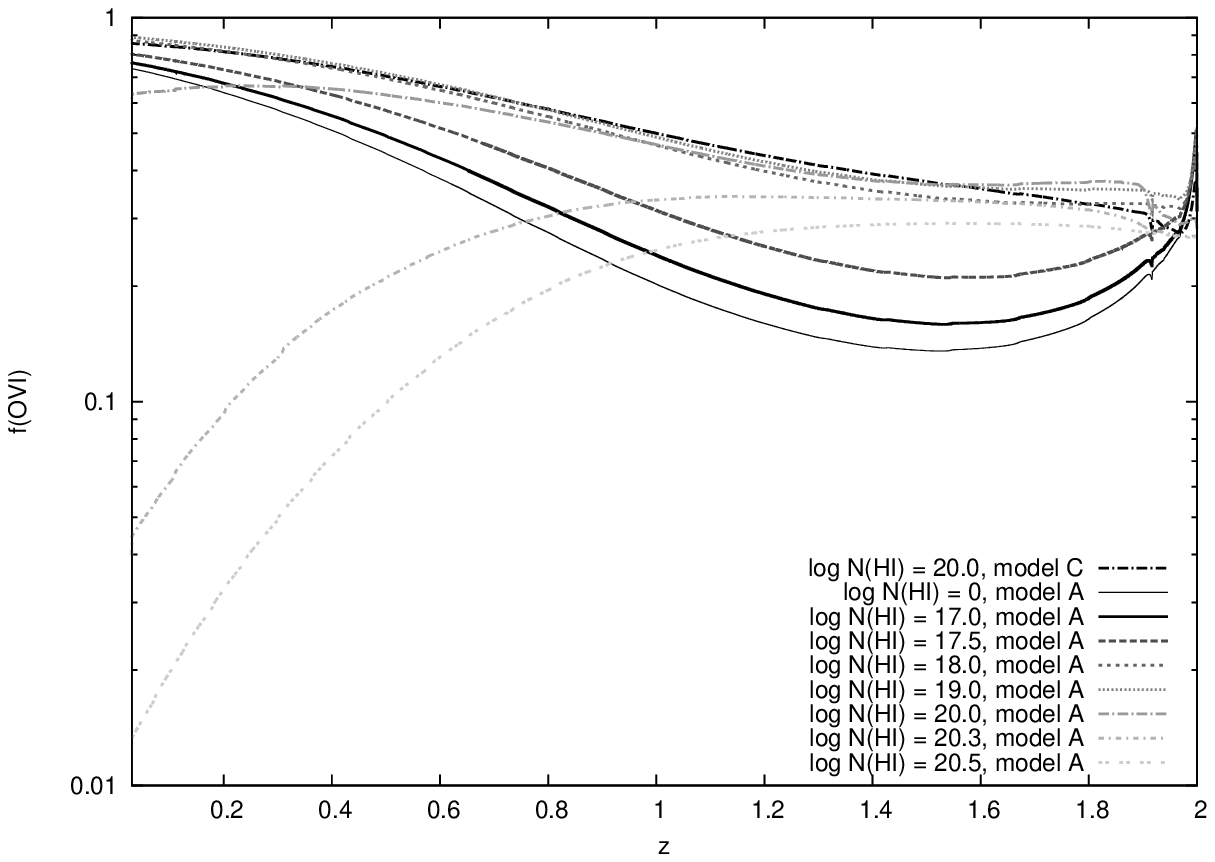}
\caption{
{  
The galactic part of the spectrum {  in model A} at distance $d=100$~kpc and redshift $z=0.1$ for several values
$N_{\rm HI}$ (upper panel), and the evolution of OVI fraction in gas exposed this spectrum (lower panel). The thick 
black line on the upper panel shows the extragalactic part of the spectrum. Redshift dependence of the fraction of 
OVI ions on the lower panel for different column densities on lower panel shows a catastrophic change at $N_{\rm HI}> 10^{20}$
cm$^{-2}$ with a heavily depressed $f({\rm OVI})$ at lower $z$. The ratio $N_{\rm HI} / N_{\rm HeI} =10$ is assumed.
}
}
\label{fig-ovi-HI}
\end{figure}

Figure~\ref{fig-ovi-HI} presents the galactic part of the spectrum at distance $d=100$~kpc and redshift $z=0.1$ for
several values $N_{\rm HI}$ (upper panel), and the evolution of OVI fraction in gas exposed to this radiation field
(lower panel). For column density $N_{\rm HI} = 10^{20}$~cm$^{-2}$ the galactic spectrum in model A is fully absorbed 
in the range 13.6-80~eV. Only for a hundredth of this column density the attenuation becomes less depressing.
A lower column density 
$N_{\rm HI} \simlt 10^{19}$~cm$^{-2}$ allows a higher fraction of ionizing photons with the energy above the 
ionization potential of OV (113~eV) to penetrate into the halo and support a higher fraction of  OVI 
(lower panel): the OVI fraction reaches $\sim 0.8-0.9$ and slightly increases 
for lower column densities. This value is considerably higher than in collisionally ionized gas
and in gas photoionized only by the extragalactic background. The reason for such a high fraction of OVI 
is in the break at the energy
slightly lower than the OVI ionization potential (136~eV). It is worth noting that the uncertainty of the spectral energy 
distribution in this energy range within stellar population codes is rather high. This partly stems from the 
stellar atmosphere models and the rarity of extremely massive stars \citep{rauch03}. On one hand this is 
a shortcoming of our model, though on the other, the rarity of massive stars and their short lifetime allows us to think 
that stellar contribution into circumgalactic ionizing field falls steeply down at energies $E\simgt 125-130$~eV 
\citep[see the spectra of the hottest stars in][]{rauch03}, which is lower than the OVI ionization potential (136~eV). 
However it is important to note that 
that the predominance of galactic flux over the extragalactic background in the range around 110-130~eV results in OVI
fraction as high as 0.6-0.8. Higher HI column density in underlying galactic discs, $N_{\rm HI} \simgt {\rm
several} \times 10^{20}$~cm$^{-2}$, heavily erases galactic flux in the range $E\sim 110-130$~eV, which, for instance 
in Model A becomes lower than the extragalactic background and the effect of an enhanced OVI fraction vanishes.

\subsection{Stability against spectral variations at $\simlt$91~\AA}

\begin{figure}
\center
\includegraphics[width=77mm]{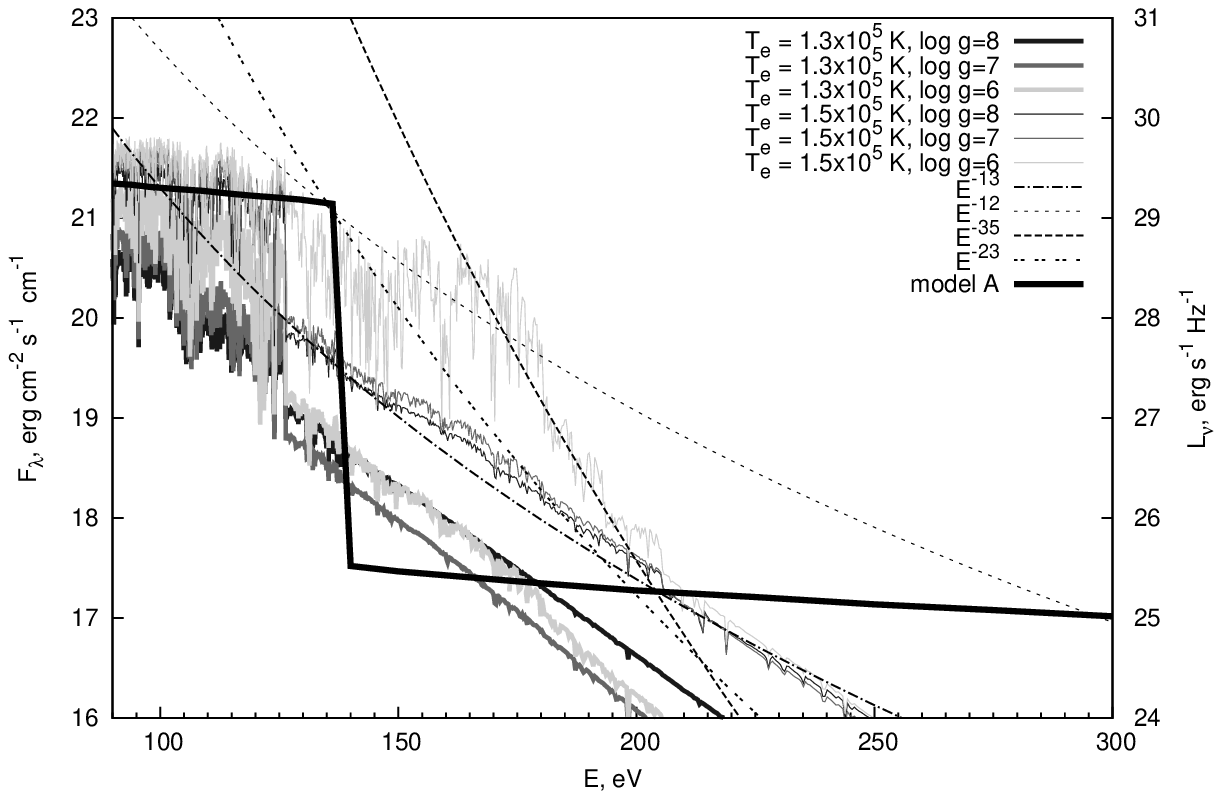}
\includegraphics[width=75mm]{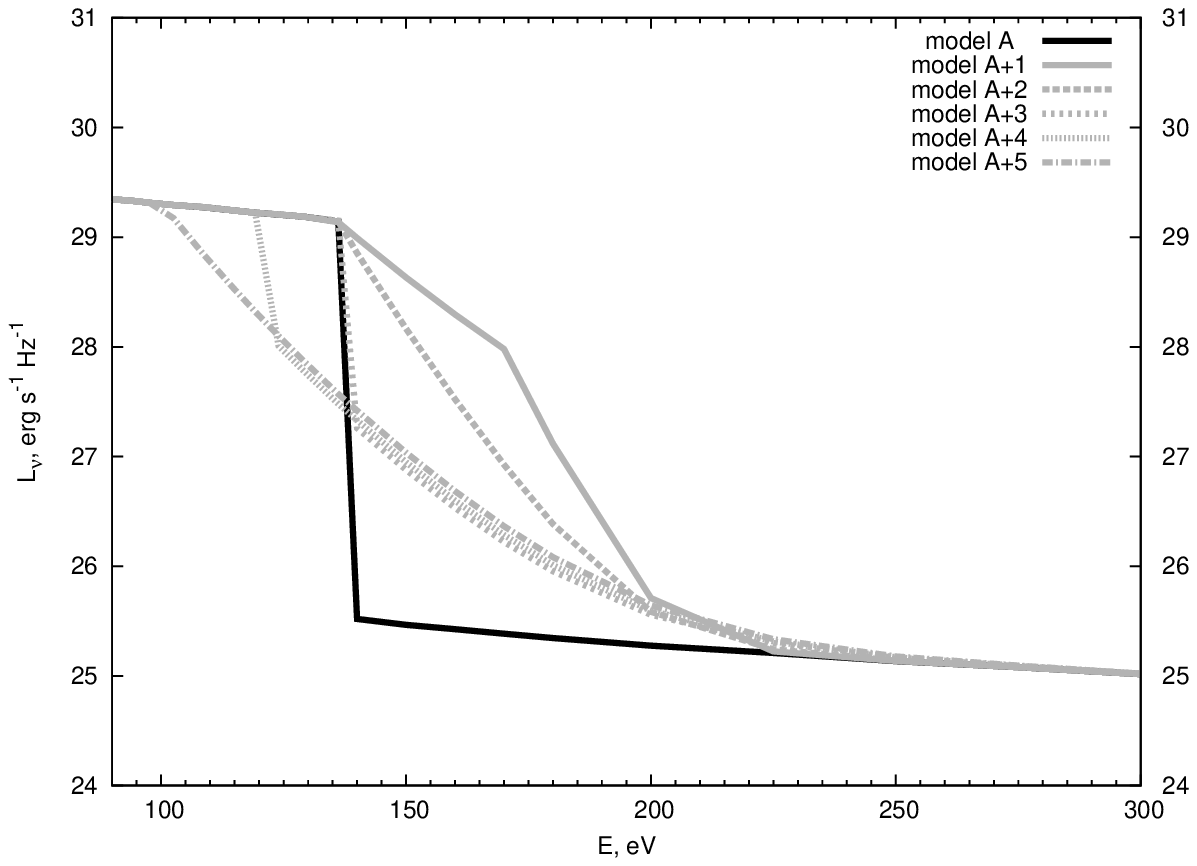}
\includegraphics[width=75mm]{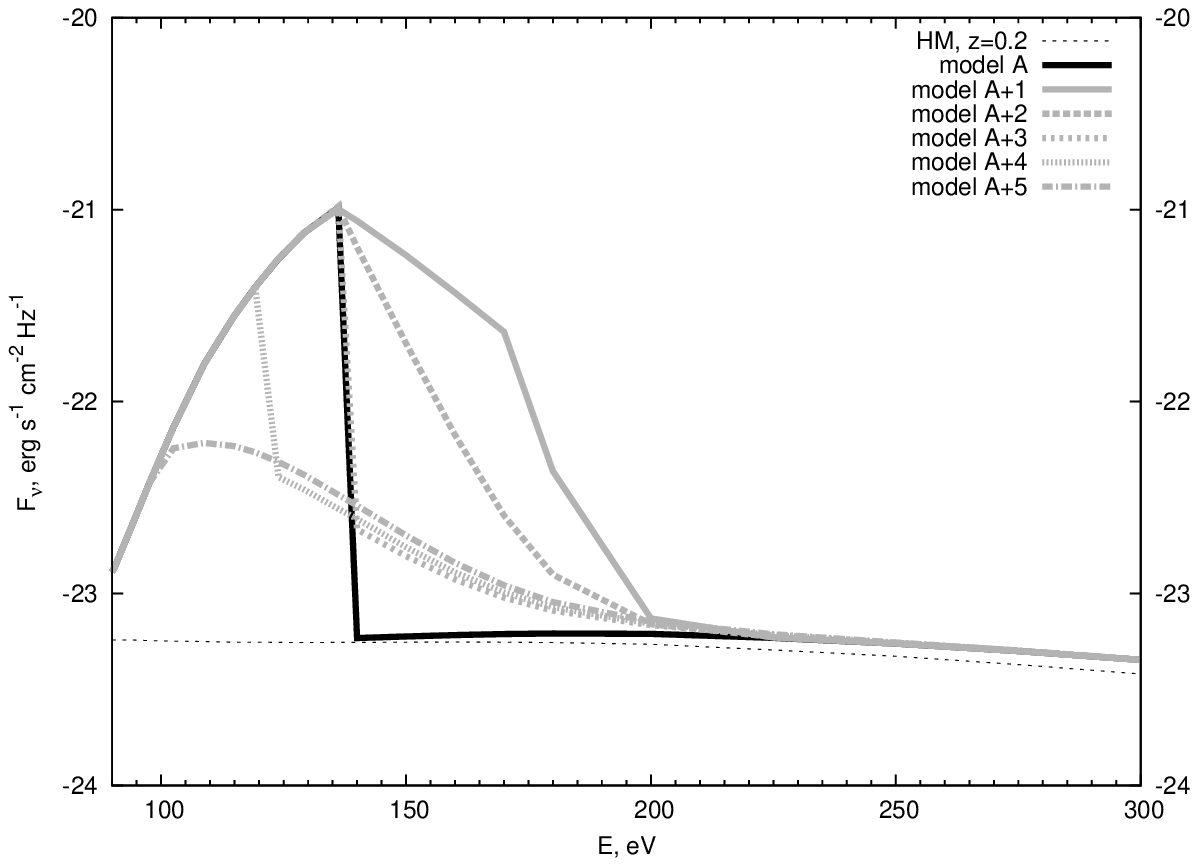}
\caption{
{ 
{\it Upper panel:} Stellar spectra for $T_e = 1.3\times 10^5$~K (thick grey) and $1.5\times 10^5$~K (thin grey lines)
for surface gravity $g =10^8$, $10^7$ and $10^6$~cm~s$^{-2}$ (dark grey, grey and light grey lines) at solar metallicity
taken from the stellar library calculated by \citet{rauch03} are shown by grey lines (the left axis). The galactic spectral
luminosity accepted in model A is depicted by thick black line (the right axis). Several power-law functions $f_\nu \sim E^{-\alpha}$ 
are shown for comparison.
{\it Middle panel:} The spectral luminosity in models A (black line) and A+1..A+5 at $z=0.2$, corresponding to different 
power-law index $\alpha$ as shown by grey lines.
{\it Bottom panel:} The cumulative ionizing background flux at a distance from the galaxy $d=100$~kpc and redshift $z=0.2$ 
in models A, A+1..A+5 (thick lines), the extragalactic ionizing background (thin dotted line). Models C+1, C+2, ... are defined as 
the model C with the spectra accepted for models A+1, A+2, ..., correspondingly. 
}
}
\label{fig-spec-z02}
\end{figure}

\begin{figure}
\center
\includegraphics[width=80mm]{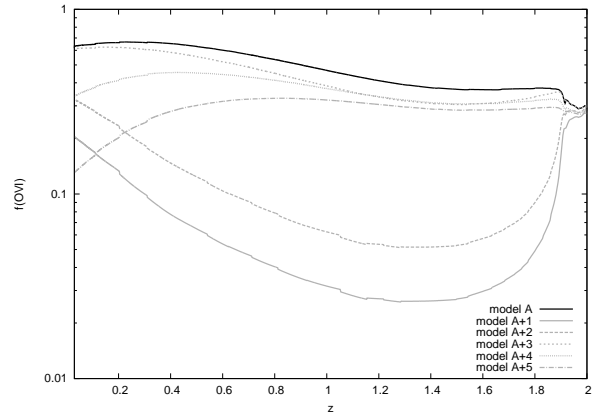}
\caption{
{ 
The OVI fraction in gas at a distance $d=100$~kpc from the galaxy exposed to the spectra corresponding to models A, A+1..A+5. 
}
}
\label{fig-ovi-spec}
\end{figure}

{
It is obviously clear from discussion in Section 3.2 that the enhanced OVI fraction is mainly due to the excess of 
galactic photons with $E>113$~eV in the bump at $E\sim 90-136$~eV. This bump maintains the OVI fraction at high level 
and keeps them against recombination to lower ionic states. The left border of this spectral feature, formed by  
to strong absorption of the galactic ionizing radiation in the disc, is irrelevant from the point of view of the 
maintainance of OVI -- even in the absence of any absorption it remains enhanced (see Figure~\ref{fig-ovi-HI}). 
The right edge of the bump is, however, artificial
in origin: as a matter of fact, the cut at 136~eV (91~\AA) is a choice of the developers of stellar population codes (e.g. PEGASE, Starburst99,
Galaxev). Therefore, in the following we discuss convergence of our results against possible galactic spectral
variations at energy higher than 136~eV.
}

{
Extreme ultraviolet and soft X-ray photons ($E\simgt 100$~eV) can be produced by post-AGB stars before entering the 
white dwarf cooling phase \citep[e.g.][]{werner}. During this very bright evolution phase lasting $\simlt 10^5$~yrs 
the effective temperature may reach more than $T_e\sim 10^5$~K, while the surface gravity varies as $g \sim 10^{5.5}$ 
to $10^{9}$~cm~s$^{-2}$. Figure~\ref{fig-spec-z02} (upper panel) presents the two spectra for $T_e = 1.3\times 10^5$~K 
and $1.5\times 10^5$~K for solar metallicity taken from the stellar library \citep{rauch03}\footnote{ This is up to 
now the most complete library which contains spectra up to 1~\AA \ for hot compact stars.}. It is seen that the amount 
of photons with $E\simgt 136$~eV increases when the surface gravity $g$ decreases from $10^7$ to $10^6$~cm~s$^{-2}$ for 
$T_e = 1.5\times 10^5$~K. On the other hand even a small decrease of effective temperature results in a shortage of 
photons with $E\simgt 125$~eV independent on $g$. The majority of observed post-AGB stars is known to have masses below 
$0.6~\msun$ and temperatures $T_e\sim 10^5$ to $1.5\times 10^5$~K \citep[e.g.][]{werner}, thus falling into the range 
shown here. For several spectra a break around $E\sim 125$~eV is clearly seen. One can think that in order to conservatively 
account this break in stellar population codes the value 136~eV (91~\AA) was choosen as an upper limit of energy in their 
stellar library. It is clearly seen though that even the spectra with a break at $E\sim 125$~eV continue up to $E\sim 150-170$~eV 
and do not show such a strong decrease of flux at $E\simgt 136$~eV as used in our calculations. One can think that this gradually 
decreasing flux above $E\sim 125$~eV produces qualitatively similar effects as does the flux with the break at $E\simgt 136$~eV.
}

{
In order to check this tentative conclusion we performed calculations of ionic composition for several models of the ionizing 
flux above 136~eV
we construct several spectra and test the appearence of enhanced OVI fraction. We smooth the spectra calculated in PEGASE
code at energies $E \sim 100-200$~eV and match them at $E \simgt 200$~eV to the spectrum obtained from the ``$L_X - SFR$''
relation as described in Section 2.3 \citep[see ][]{gilfanov04}. We proceed in the manner that the resulting spectrum would be qualitatively close 
to the spectra depicted in the upper panel. Figure~\ref{fig-spec-z02} (middle panel) shows these spectra normalized on the
spectral luminosity in model A.} 
{ We calculate the ionization and thermal evolution of gas exposed to a cumulative spectrum which includes the extragalactic 
and the galactic radiation field.  Figure~\ref{fig-spec-z02} (bottom panel)
presents an example of the cumulative ionizing background flux at a distance $d=100$~kpc from the galaxy in models A, 
A+1..A+5 located  at a redshift $z=0.2$ (thick lines). Figure~\ref{fig-ovi-spec} shows evolution of OVI fraction in gas exposed to the ionizing flux as  
in models A, A+1..A+5. It is readily seen that in all A+ models the OVI fraction is greater than 0.1 at redshifts $z\simlt 0.2$.  
The OVI fraction is several times lower than in model A, but this is still obviously higher than $\simlt 0.03$, which 
establishes in gas exposed only to the extragalactic radiation (compare to the dot-dot line for log~N(HI)=20.5 in lower panel 
of Figure~\ref{fig-ovi-HI}). In models A+1 and A+2 oxygen at $z\simgt 1$ is mainly locked in OVII ions. Starting from 
$z\sim 1$ the ionization production rate of OVII ions decreases such that recombinations become efficient to replenish lower 
ionization states, and as a result the fraction of OVI increases at $z<1$ substantially: from $\sim 0.03$ at $z=1.5$ 
to $0.15$ in A+1 and $\simgt 0.3$ in A+2 at $z\simlt 0.2$. This balance between OVII and OVI is kept then quasi-steady 
down to $z\sim 0$. At large distances the contribution from galactic photons decreases resulting in a higher fraction of 
OVI in models A+1 and A+2. Note that contrary to models A+1 and A+2 the flux in model A+5 is low and oxygen accumulates 
mostly in OIV-OV states. 
}

{
Higher galactic luminosities presented in a set of models C and C+3..C+5, result in a higher OVI fraction: $f({\rm OVI}) \simgt 0.5$ in 
gas located at $d=100$~kpc at $z\simlt 0.2$. However, in models C+1 and C+2 the ionization flux at $E\simgt 125$ eV is excessive 
and the fraction of OVI decreases to 0.03 and 0.07 at $z=0.2$, 
and reaches only 0.05 and 0.1, at $z=0$. At larger distances from a galaxy where 
the flux drops models C+1 and C+2 show an increase of OVI fraction: e.g., in model C+1 it increases to $\sim 0.07$ (0.11) at 150~kpc and $\sim 0.1$ (0.17) at 200~kpc at $z=0.2$ ($z=0$). 
}

{
It is readily seen thus that in the majority of these models OVI fraction remains high and reaches $\simgt 0.2$ at
distance $d\simgt 100$~kpc from a galaxy located at $z\simlt 0.2$. This is more than an order of magnitude higher than can be  
reached in gas exposed only to the extragalactic radiation at $z\simlt 0.2$. In model C+1 the OVI fraction
reaches 0.1 only for $d\simgt 150$~kpc ($z\simlt 0.2$), while at closer distances most of oxygen is confined into OVII. 
It is worth stressing though that the model C+1 
can be considered as a model with an extreme possible contribution from post-AGB stars, in the sense that models with 
harder spectrum are unlikely \citep[e.g.][]{werner}, thus one can expect that within conservative models of galactic X-ray 
spectrum the effect of enhanced OVI remains stable.
}

\subsection{OVI in galactic haloes}

In the previous section we have found that OVI fraction can reach high values of $f({\rm OVI}) \sim 0.4-0.9$ under the
action of ionizing radiation from the underlying galactic stellar population along with the extragalactic radiation 
field. Such OVI fraction is at least an order higher than the maximum $\sim 0.1$ reached in commonly used models
with gas ionized collisionally and/or by the extragalactic radiation. Assuming the spherical symmetry
and using the time-dependent OVI radial distribution around a host galaxy one can find the OVI column densities
along the line of sight crossing the galactic halo at impact parameter~$b$. {  We integrate from $r = 50$~kpc to
300~kpc.}

\begin{figure}
\center
\includegraphics[width=80mm]{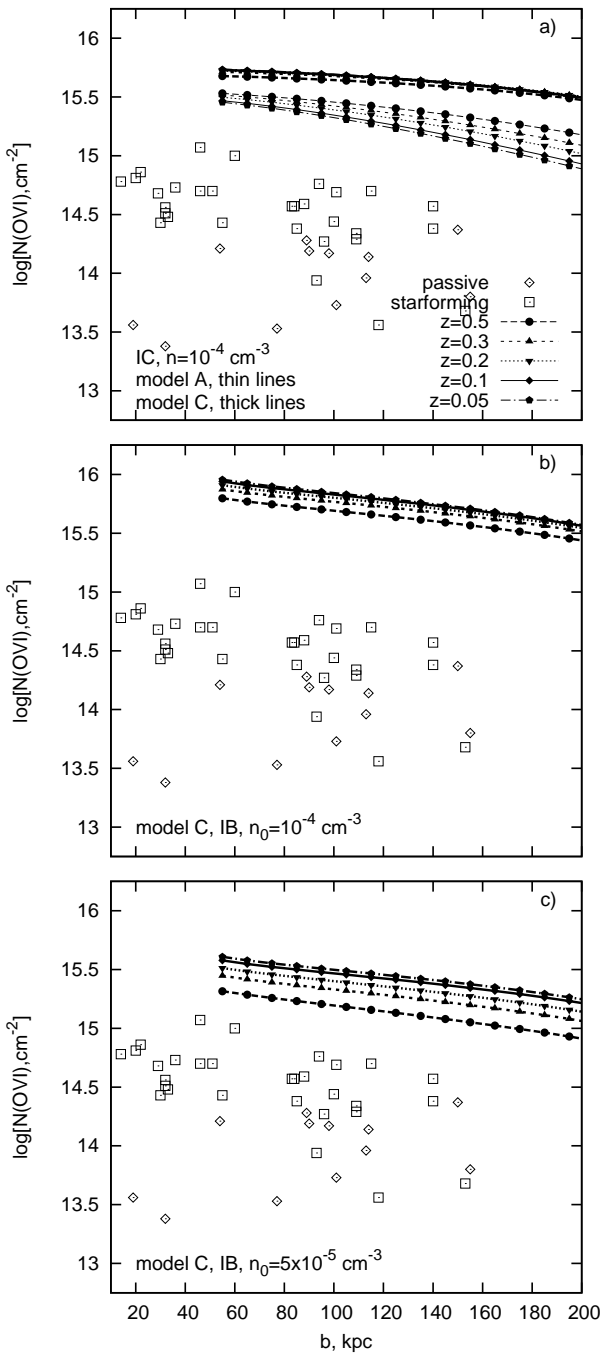}
\caption{
The dependence of the OVI column density on the impact parameter. 
Thin and thick lines correspond to models A and C (small symbols), respectively. 
Lines with filled symbols present the dependence for
different redshift (see details on the upper panel). The upper panel depicts the column density for isochoric (IC)
gas with volumetric density $n=10^{-4}$ cm$^{-3}$, the middle and lower panels show the column density for the isobaric 
(IB) regime with the initial density $10^{-4}$~cm$^{-3}$ and $5\times 10^{-5}$~cm$^{-3}$, respectively; 
the metallicity is $0.1~\zsun$. Open rhombs and squares depict the observed column densities in the passive
and starforming galaxies, correspondingly \citep{tumlinson11}; {  integration is over $r = 50$~kpc to
300~kpc, different symbols correspond to different redshifts.}
}
\label{figcd_ovi}
\end{figure}

\begin{figure}
\center
\includegraphics[width=80mm]{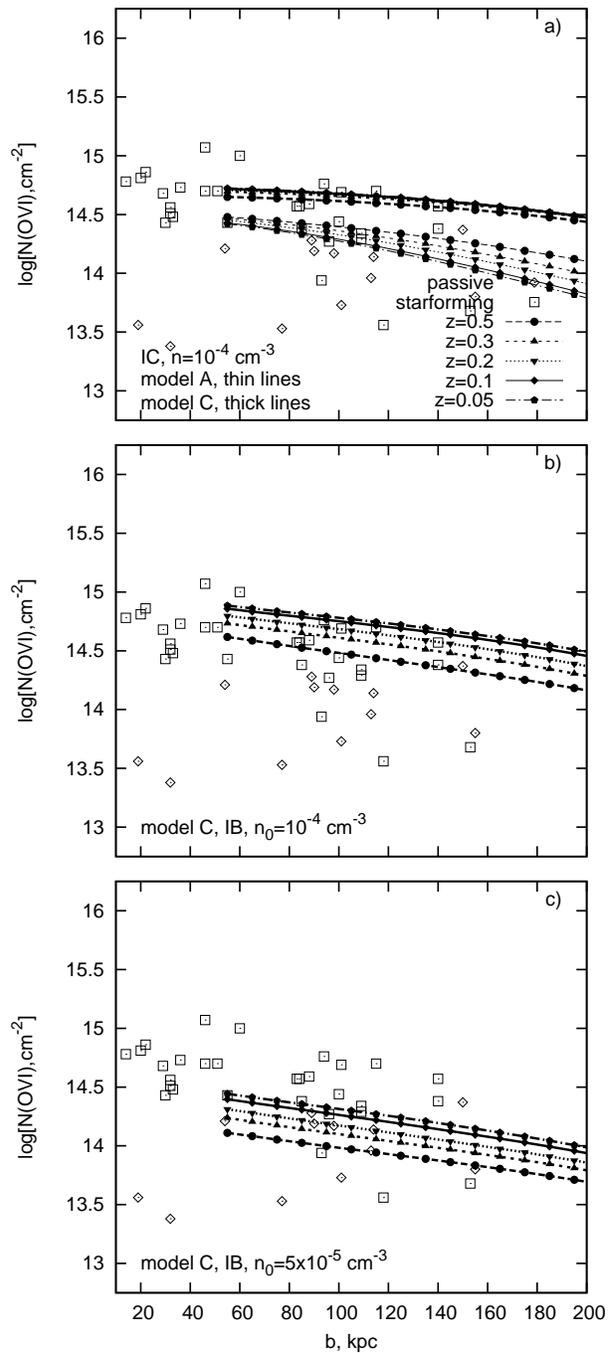}
\caption{
The same as in Figure~\ref{figcd_ovi}, but for gas metallicity $0.01~\zsun$. 
}
\label{figcd_ovi2}
\end{figure}

{  
Figure~\ref{figcd_ovi} (upper panel) presents the dependence of $N({\rm OVI})$ on impact parameter 
$b$ in a $Z=0.1\zsun$ isochoric gas within the models A and C. The OVI column density 
ranges from $N({\rm OVI})\sim 10^{14.9}$ to $\sim 10^{ 15.7}$~cm$^{-2}$ at the 
impact parameter $b \simlt 200$~kpc -- a factor of 3-10 higher than observed by \citet{tumlinson11}.
Therefore circumgalactic gas of an order of magnitude lower metallicity $Z=0.01\zsun$ would provide 
good agreement with observations as shown in Figure~\ref{figcd_ovi2} (upper panel).  
It is worth noting though that spatial distribution of metals in circumgalactic environment 
is highly nonhomogeneous \citep[e.g.][]{simcoe06,mix04}, and it is most likely for metals to 
be locked in clumps of smaller covering factor resulting in a proportional decrease of the column density. 
}

{  
Figure~\ref{figcd_ovi} (middle and lower panels) show the column density of OVI in isobaric gas with the initial 
density $10^{-4}$ and $5\times 10^{-5}$~cm$^{-3}$, respectively. Isobaric regime is supported by radiation losses 
and heating from ionizing radiation. Photo-heating grows at large distances from the galaxy, $r\simgt 250$~kpc where
contribution from OVI ions to cooling decreases. At lower distances OVI fraction increases and it becomes a dominant 
coolant resulting in a rather efficient cooling. Consequently, column densities grow under isobaric compression:
$N({\rm OVI})$ can reach $\sim 10^{15.5 - 15.9}$~cm$^{-2}$ for the initial density $10^{-4}$~cm$^{-3}$, and $\sim 
10^{15 - 15.6}$~cm$^{-2}$ for $5\times 10^{-5}$~cm$^{-3}$. Similarly to the isochoric case gas with lower metallicity
(0.01~$\zsun$) fits observational data better.  
}

{
Such high column densities are reached for galactic spectra with a cut at 91~\AA \ (see footnote 
in Section 3.1). In reality though galactic spectra do not show such sharp breaks. In Section 
3.3 we considered galactic spectra approximated smoothly around $\sim 100-300$~eV to account possible variations of 
contribution from post-AGB stars. We have found that in the low redshift range, $z\simlt 0.2$, OVI fraction in gas exposed 
to such approximated spectra  
remains sufficiently high in the majority of models considered here.  In all cases we have a considerable (an order of magnitude) 
excess of the column densities
caluculated in models A, C and with respect to the observed ones as seen in Figure~\ref{figcd_ovi}).
}

\citet{tumlinson11} have argued that the circumgalactic medium can be a significant reservoir of the ejected 
material from galaxies. In order to explain large column densities of OVI observed in the haloes they have had  
to assume that the circumgalactic gas has nearly solar metallicity. This assumption meets though difficulties 
because the minimum oxygen mass in the halo obtained within such an assumption reaches around 10-70\% of the 
total oxygen mass in the ISM \citep{tumlinson11}, which requires in turn unprecedentedly high mass exchange 
between the galactic ISM disk and a huge circumgalactic reservoir extending up to 150~kpc. Higher OVI fraction
obtained in our model reduces the estimates to a more reasonable level, and as a consequence weakens 
constrains on the sources of oxygen in underlying galaxies. 

Contrary, an order of magnitude lower metallicity results in a more consistent scenario of metal production and 
mass exchange between galactic and circumgalactic gas. Indeed, the total mass of the circumgalactic gas in OVI 
haloes within $\simlt 300$~kpc is about $(3-4)\times 10^{10}~\msun$ independent on redshift as estimated from HI 
data \citep{HIprochaska,HIsteidel}. If such gas would be enriched to the metallicity as high as $0.1~\zsun$,
then the oxygen mass in it equals $\sim 2\times 10^7~\msun$. This amount can be produced during a 0.2~Gyr period 
in a galaxy with SFR $\sim 6~\msun/yr$ (as in our model A), with 0.015~$\msun$ of oxygen returned to the ISM per 
each 1~$\msun$ star formed \citep[e.g., ][]{yields-thomas}. This timescale is significantly shorter than the 
whole period of the evolution -- several Gyrs, and comparable to the initial period of evolution in our models 
with the SFR kept nearly constant {  (upper panel on Figure~\ref{figssfr-time})}. Because of the high SFR during initial 
several hundreds Myrs, gas can be ejected from the disk with velocities $\simgt 100$~km/s sufficient to reach
$\sim$100~kpc in approximately 1~Gyr. {  We assumed throughout that metals are honogeneously distributed in a spherical 
layer between 50 and 300~kpc. This can be an overestimate, because in general the volume occupied by ejected metals 
depends on the galaxy mass and for dwarf galaxies can be smaller \citep[e.g.][]{mix00}. Moreover, metals are locked 
most likely into small-size dense fragments \citep[e.g.][]{simcoe06}, where radiative cooling is efficient. It can 
result in a rapid cooling and compression and make such fragments to escape detection.} Dwarf satellites may add 
metals in the halo, though their contribution is apparently small because of low metallicity in their ISM:  
$<0.1~\zsun$ \citep[e.g.,][]{salvadori09,ryabova11} and massive dwarfs are rare \citep[e.g.,][]{koposov08}.

\subsection{{  OVI in galactic haloes: photoequilibrium}}

{  
In extended galactic haloes gas with a low density, $\sim 10^{-5}-10^{-3}$~cm$^{-3}$, is exposed to rather a high
ionizing background consisting of galactic and extragalactic components. Under such conditions the ionic state 
may reach photoequilibrium\footnote{{  The analysis, when collisions become significant compared to photoionization,
is out of scope of this paper, details can be found in \citep{v11}}}. Ionic composition under  photoequilibrium
is calculated with making use the CLOUDY. We assume the time-dependent ionizing background described in Section 2.3.
The difference between our calculations and the ones performed by \citet{tumlinson11}, is that besides the 
Haardt-Madau extragalactic radiation field used in \citep{tumlinson11} we add the galactic contribution, which as we 
showed above may have crucial consequences: the excess of photons with $E\sim 13.6-130$~eV
from the underlying stellar population competes the extragalactic radiation,
while X-ray photons in the excess with $E>113$~eV substantially enhance OVI fraction (Section
3.2). 
}

{  
For the fiducial column densities $N_{{\rm HI}} = 10^{20}$~cm$^{-2}$ and $N_{{\rm HeI}} = 10^{19}$~cm$^{-2}$  
OVI fraction in photoequilibrium is several magnitudes smaller than in our time-dependent models. 
In photoequilibrium oxygen is mainly locked in less ionized states, OIV-OV, and OVI column density drops to
$10^{12.5-13}$~cm$^{-2}$. When $N_{{\rm HI}}$ and $N_{{\rm HeI}}$ decrease 
the fraction of OVI instead increases. 
Even for $N_{{\rm HI}}=5\times10^{19}$~cm$^{-2}$ (here $N_{\rm HI} 
/ N_{\rm HeI} =10$ is assumed) the OVI column density reaches $\sim 10^{14.2-14.7}$~cm$^{-2}$ at the impact distances 
$b=50-200$~kpc 
and $z=0.1$. Further decrease of $N_{{\rm HI}}$ results in OVI fraction to grow up to $\sim 0.4-0.8$, 
and consequently in higher OVI 
column densities. 
}

\begin{figure}
\center
\includegraphics[width=80mm]{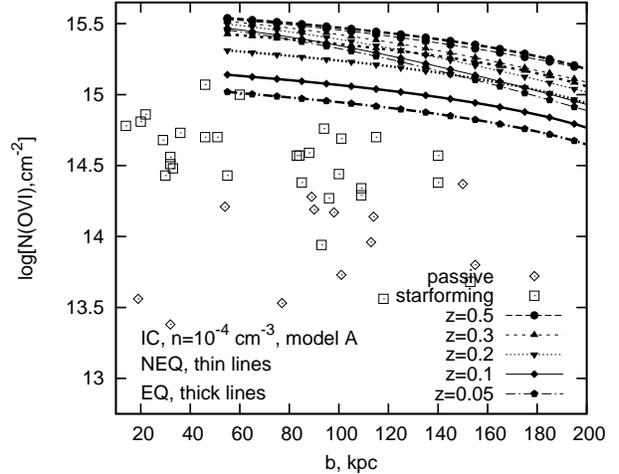}
\caption{
{  
The dependence of the OVI column density on the impact parameter for isochoric gas exposed to the ionizing spectrum 
in model A. In the photoequilibrium $N_{{\rm HI}}=3 \times 10^{19}$~cm$^{-2}$, whereas in time-dependent 
model $N_{{\rm HI}}=10^{20}$~cm$^{-2}$ is adopted ($N_{\rm
HI} / N_{\rm HeI} =10$ is assumed). 
Thin and thick lines correspond to nonequilibrium and photoequilibrium, respectively;
density and metallicity are $n=10^{-4}$ cm$^{-3}$ and $0.1~\zsun$, 
the other notations are as in Figure~\ref{figcd_ovi}.
}
}
\label{figcd_ovi-eq}
\end{figure}

{  
Figure~\ref{figcd_ovi-eq} presents the dependence of the OVI column density on the impact parameter 
for photoequilibrium model with $N_{{\rm HI}}=3 \times 10^{19}$~cm$^{-2}$ and time-dependent model with $N_{{\rm
HI}}=10^{20}$~cm$^{-2}$. The clearly seen considerable difference between photoequilibrium and time-dependent 
models originates 
from a well-known fact that gas under nonequilibrium conditions is overionized with respect to 
what occurs under equilibrium \citep[e.g.][]{gs07,ss84,v11}. When $N_{{\rm HI}}$ decreases, the absorption of 
galactic ionizing radiation falls down and the peak in the spectrum around $\sim 80-130$~eV grows. 
At such conditions photoionization time shortens,
$t_p\sim h\nu /(F_\nu \sigma_\nu \Delta\nu)$, and the ionic composition approaches photoequilibrium. At low $z$
the extragalactic flux decreases: for example, for $z=0.1$ $F_\nu \sim 3\times 10^{-24}$~erg~s$^{-1}$~cm$^{-2}$~Hz$^{-1}$
at $\sim 100$~eV (Figure~\ref{fig-ovi-HI}). As a result the phoionization timescale $t_p$ of OVI ions increases up to 
several hundreds Myr, 
here $\sigma_\nu \sim 10^{-18}$~cm$^{-2}$ and $\nu/\Delta\nu \sim 10$ are assumed; $\nu=113.9$~eV is the OVI ionization
potential and $\Delta\nu$ is the width of the peak between 113-136~eV. This is about several times shorter than 
the recombination time for gas with $n=10^{-4}$~cm$^{-3}$ making nonequilibrium effects important. 
}

{  
Smaller absorption in the galactic disc provides higher galactic ionizing flux} {  and shorter   
photoionization timescale} {   such that ionic composition shifts to photoequilibrium. Thus, nonequilibrium effects are
important for OVI ionization kinetics in the circumgalactic gas with $n \simgt 10^{-4}$~cm$^{-3}$ at $z\simlt 0.5$. 
t is therefore can be concluded that the main factor which provides high OVI column densities in massive starforming 
galaxies at low and moderate
metallicities is the excess of photons with energies 113-130~eV from the stellar population of a host galaxy. 
}

\section{Conclusions}

\noindent

In this paper we have presented the {  photoequilibrium} and nonequilibrium (time-dependent) ionization and 
thermal {  state} of circumgalactic gas located at distances up to $\sim 50-300$~kpc around starforming galaxies, 
and exposed to both extragalactic and galactic time-dependent ionizing background. For the extragalactic background 
we considered the spectra obtained by \citet{hm01}. Using the PEGASE code \citep{pegase97} we have calculated chemical
and spectro-photometric evolution of galaxies, and have chosen the two models, whose specific star formation rate 
(${\rm sSFR} = {\rm SFR}/M_*$) and stellar masses are close to the starforming galaxies with large OVI column densities
observed in \citet{tumlinson11}. 

We have found that
\begin{itemize}
 \item the maximum OVI fraction can reach $\sim 0.4-0.9$ under physical conditions (gas density and metallicity, and the
 spectrum shape), which are typical in haloes of starforming galaxies; such a high OVI fraction is due to
 galactic photons with $E\simgt 113$~eV; { the effect of enhanced OVI remains stable within conservative models of 
 galactic X-ray spectrum fluctuations at $E\sim 100-200$~eV};
 \item due to such high fraction of OVI its column density ranges in ${\rm N(OVI)}\sim 10^{14.9 - 15.7}$~cm$^{-2}$ even 
 for a low  metallicity $Z = 0.1\zsun$, and $\sim 10^{14 - 15}$~cm$^{-2}$ for $Z= 0.01\zsun$ at impact parameters 
 up to $\simlt 200$~kpc; this results in several times more conservative estimate of the oxygen mass in haloes 
 compared to $M_O = 1.2\times 10^7 (0.2/f_{OVI})~\msun$ \citep{tumlinson11}.
 \item we have shown therefore that the large OVI column densities observed in haloes of starforming galaxies
 \citep{tumlinson11} can be found in circumgalactic conditions with nearly 0.01-0.1 of solar metallicity, and 
 correspondingly the requirements to the sources of oxygen in the extended haloes become reasonably conservative.
\end{itemize}
{  
High OVI column densities in haloes of starforming galaxies can emerge under  
photoequilibrium  as well as under nonequilibrium conditions. The main source is a high radiation flux of photons  
with $E\simgt 113$~eV from the stellar population of starforming galaxies. Nonequilibrium 
effects for OVI ionization kinetics are important in the circumgalactic environment 
with $n \simgt 10^{-4}$~cm$^{-3}$ at $z\simlt 0.5$.
}

{  
Very recently \citet{lehner-ovi} have reported about high OVI column densities in the circumgalactic medium 
at $z\sim 2-3$ with $N({\rm OVI})$ reaching $\sim 10^{15}$~cm$^{-2}$; the observational sample 
includes absorbers of different type: from Lyman limit to damped Ly$\alpha$ systems. From our point of view 
such a high OVI column density can be due to the excess of galactic ionizing photons with $E\sim 113-135$~eV 
over the extragalactic background, while the metallicity might be rather low.
}

\section{Acknowledgements}

\noindent

We thank Jason Tumlinson for 
providing data and discussion, and {  the anonymous referee for valuable comments and pointing to a mistake}. 
This work is supported by the RFBR through the grants 12-02-00365, 12-02-00917, 12-02-92704, 
and by the Russian Federal Task Program "Research and operations on priority directions of 
development of the science and technology complex of Russia for 2009-2013" (state contracts 
14.A18.21.1304, 2.5641.2011 and 14.A18.21.1179). EV is grateful for support from the "Dynasty" 
foundation.


\end{document}